\documentstyle[emulate_apj,apjfonts,epsf]{article}

\gdef\h50min{$h_{50}^{-1}$}
\gdef\1054{MS\,1054$-$03}
\gdef\kms{km\,s$^{-1}$}
\gdef\3727{O\,[{\sc ii}]\,3727\,\AA}
\lefthead{van Dokkum et al.}
\righthead{Mergers in \1054{} at $z=0.83$}
\slugcomment{Accepted for publication in ApJ Letters}
\begin{document}
\title{A High Merger Fraction in the Rich Cluster \1054{} at $z=0.83$:
Direct Evidence for Hierarchical Formation of Massive
Galaxies\altaffilmark{1,2}}

\author{
Pieter G. van Dokkum\altaffilmark{3,4},
Marijn Franx\altaffilmark{4}, Daniel Fabricant\altaffilmark{5},
Daniel D. Kelson\altaffilmark{6},
and Garth D. Illingworth\altaffilmark{7}
}

\begin{abstract}

We present a morphological study of the galaxy population of the
luminous X-ray cluster \1054{} at $z=0.83$. The sample consists of 81
spectroscopically confirmed cluster members in a $3 \times
2$\,\h50min\,Mpc area imaged in F606W and F814W with WFPC2.  We find
thirteen ongoing mergers in \1054, comprising $17$\,\% of the $L \gtrsim
L_*$ cluster population. Most of these mergers will likely evolve into
luminous ($\sim 2 L_*$) elliptical galaxies, and some may evolve into S0
galaxies. Assuming the galaxy population in \1054{}
is typical for its redshift
it is estimated that $\sim 50$\,\% of present-day cluster
ellipticals experienced a major merger at $z < 1$. 
The mergers are
preferentially found in the outskirts of the cluster, and probably occur
in small infalling clumps.  Morphologies, spectra, and colors of the
mergers show that their progenitors were typically E/S0s or early-type
spirals with mean stellar formation redshifts $z_* \gtrsim 1.7$. The red
colors of the merger remnants are consistent with the low scatter in the
color-magnitude relation in rich clusters at lower redshift.  The
discovery of a high fraction of mergers in this young cluster is direct
evidence against formation of ellipticals in a single ``monolithic''
collapse at high redshift, and in qualitative agreement with predictions
of hierarchical models for structure formation.

\end{abstract}

\section{Introduction}

\altaffiltext{1}
{Based on observations with the NASA/ESA {\em Hubble Space
Telescope}, obtained at the Space Telescope Science Institute, which
is operated by AURA, Inc., under NASA contract NAS 5--26555.}
\altaffiltext{2}
{Based on observations obtained at the W.\ M.\ Keck Observatory,
which is operated jointly by the Californian Institute of
Technology and the University of California.}
\altaffiltext{3}
{Kapteyn Astronomical Institute, P.O. Box 800, NL-9700 AV,
Groningen, The Netherlands}
\altaffiltext{4}
{Leiden Observatory, P.O. Box 9513, NL-2300 RA, Leiden, The Netherlands}
\altaffiltext{5}
{Harvard-Smithsonian Center for Astrophysics, 60 Garden Street,
Cambridge, MA 02318}
\altaffiltext{6}
{Department of Terrestrial Magnetism, Carnegie Institution of
Washington, 5241 Broad Branch Road, NW, Washington D.C., 20015}
\altaffiltext{7}
{University of California Observatories\,/\,Lick Observatory,
University of California, Santa Cruz, CA 95064}

We do not know how and when luminous elliptical galaxies were assembled.
Traditional models assume that all ellipticals are $10^{10}$
years old, having
experienced very little mass evolution after their
initial collapse (e.g., Searle, Sargent, \& Bagnuolo 1973).  In
contrast, galaxy formation models in cold dark matter (CDM) cosmologies
predict that massive galaxies were assembled in many generations of
mergers, and ellipticals experienced their last major mergers at
$z<1$ (Kauffmann 1996; Baugh, Cole, \& Frenk 1996). 

Ideally, the evolution of the mass function is determined directly from
mass measurements.  An alternative approach is to
determine the evolution of the merger fraction with redshift. 
Mergers in present-day rich clusters are rare because the probability of a
low velocity encounter is small.  If cluster ellipticals
formed in mergers these must have occurred before or
during the initial collapse of the cluster (Roos \& Aarseth 1982; Merritt
1984). 

Morphological studies of intermediate redshift clusters have shown that
mergers were indeed more common at earlier times
(Lavery \& Henry 1988 [LH88]; Lavery, Pierce, \& McClure 1992 [LPM92];
Dressler et al.\ 1994 [D94]; Couch et al.\ 1998 [C98]).
These studies indicate merger fractions of $\sim 5$\,\% in clusters
at $0.2 < z < 0.4$. However, since these mergers are
generally blue and of low luminosity they are probably not very
massive (e.g., C98).  Furthermore, Dressler et al.\ (1997)
suggest that disturbed galaxies in intermediate redshift clusters are
generally disrupted disks rather than major mergers.  These results
indicate that massive cluster ellipticals were assembled at even
higher redshift.

Recently it has become possible to extend morphological studies of
rich clusters to $z \approx 1$ (e.g., Lubin et al.\ 1998).
In this {\em Letter}, we
present results from a large area survey of the X-ray selected cluster
\1054{} at $z=0.83$.  We determine the merger fraction using an $I$
selected sample of 81 confirmed cluster members covered by a large HST
WFPC2 mosaic.  We assume $H_0 = 50$\,\kms\,Mpc$^{-1}$,
$\Omega_m = 0.3$, and $\Omega_{\Lambda} = 0$.

\section{Observations}

The \1054{} field was observed with LRIS (Oke et al.\ 1995) on the 10m
Keck II telescope. Objects were selected on the basis of their $I$ band
magnitude in a $3\arcsec$ diameter aperture ($20.0 < I < 22.7$).
These magnitudes were measured from a 900\,s Keck image, with $1\arcsec$
seeing. Spectra were taken through six multi-slit masks on 1998 February
28 and March 1. The slit masks were designed to maximize the observed
number of galaxies with $20.0 < I < 22.2$ in the area covered by our HST
imaging. Typical exposure times were 2400\,s.
A cross-correlation program was used to determine
redshifts for 186 galaxies; 80 are
cluster members. Within the area covered by our HST imaging the
completeness of our sample is 73\,\% to $I=22.2$.
The new redshifts were combined with samples of Donahue et
al.\ (1998) and Tran et al.\ (1999) giving a total of 89 confirmed
cluster members.

We obtained a large HST WFPC2 mosaic of \1054, consisting of six
independent pointings in two filters (F606W and F814W). Integration
times were 6\,500\,s in each passband and at each position. 81
confirmed cluster members are located within the HST
mosaic. Accurate magnitudes and colors were determined from the WFPC2
images. Morphological classifications of the spectroscopically confirmed
members were performed by PvD, MF, and DF,
using the F814W images (restframe $\sim B$). The classification methods
and details of our survey
are described in van Dokkum (1999).
We note that it is difficult to distinguish ellipticals from
S0s, even at low redshift (e.g., J\o{}rgensen \& Franx 1994).

\section{Mergers at $z=0.83$}

The most surprising result of our survey is the high fraction
of galaxies classified as ``merger/peculiar''.  We classified $17$\,\%
as mergers, compared to $22$\,\% ellipticals, $22$\,\% S0s, and $39$\,\%
spirals.
Merging galaxies are counted as one object.
Images of the mergers are displayed in Fig.\ 1.
The mergers
display a variety of features: double nuclei (e.g., 997), tidal tails
(e.g., 1760), and interacting doubles with distorted morphologies (e.g.,
1340).
We emphasize that all classified galaxies, including the
mergers, are spectroscopically confirmed cluster members.

The merger fraction of 17\,\% is significantly higher than in
lower redshift clusters, as discussed in the Introduction. Furthermore,
as demonstrated in Fig.\ 1a the
mergers extend to high luminosities:
five of the sixteen brightest galaxies were classified as mergers. 
Only seven out of sixteen
were classified as ellipticals or S0s, two of which (1584 and
710) have a luminous companion within 20\,\h50min\,kpc. 
The median luminosity of the mergers
is $M_B^T \approx -22$ ($\sim 2L_*$ at $z=0.83$).

The majority of the mergers will probably evolve into elliptical
galaxies (e.g., Toomre \& Toomre 1972, Barnes 1998), thereby increasing
the number fraction of massive ellipticals at later times.
However, not all mergers form ellipticals. Simulations indicate that
disks can survive mergers between galaxies of very different mass
(Barnes 1998), and so some mergers may evolve into S0s (e.g., galaxy
1583).  Most mergers show no evidence for the presence of gas to
form a new disk: only 15\,\% of the mergers have EW \3727{}\,$>5$\,\AA. 

The merger fraction is remarkably high, given the short timescale for
the merging process ($< 1$\,Gyr; e.g., Rix \& White 1989).  From the
fact that the merger fraction is similar to the elliptical fraction
we estimate that $\sim 50$\,\% of present-day cluster
ellipticals (and probably a small fraction of S0s) were assembled in
mergers at $z<1$.  This implies that a significant fraction of cluster
ellipticals have undergone major structural
change at $z<1$. 

A possible concern is that some of the mergers are misclassified.
Several show two galaxies separated by $\sim 1\arcsec$
(10\,\h50min\,kpc) in a common envelope without obvious tidal features
(e.g., galaxies 997 and 1163). One could argue that these objects are
not bound, but chance projections along the line of sight.
We have calculated the probability of chance projections from   
simulated galaxy distributions, by randomizing the position
angles of galaxies with respect to the BCG and North. The expected number of
galaxy pairs with separations $<1\arcsec$ is smaller than one.
Furthermore, both galaxies of most merging pairs were placed in
the $1\farcs 2$ wide slits. No double peaks were seen in the
cross-correlation functions, implying $\Delta v < 500$\,\kms{}
for the brightest merging pairs.  Both arguments are strong evidence
that the galaxies are bound, and that we are witnessing mergers in progress.

\section{Mechanism}

The merger fraction in \1054{} is surprisingly high, given the high
velocity dispersion of the cluster ($\approx 1170$\,\kms; Tran et al.\
1999). Since the progenitors of the mergers must have had low
relative velocities, the mergers are probably taking place in cold
subclumps which are falling into the cluster.
This is supported by the spatial distribution of the mergers, shown in
Fig.\ \ref{spatial.plot}.  The mergers occur preferentially
in the outskirts of the cluster, consistent with recent infall.
We note that the spatial distribution of the mergers is strong
aposteriori evidence against chance projections.

\setcounter{figure}{1}
\vbox{
\begin{center}
\leavevmode
\hbox{%
\epsfxsize=8cm
\epsffile{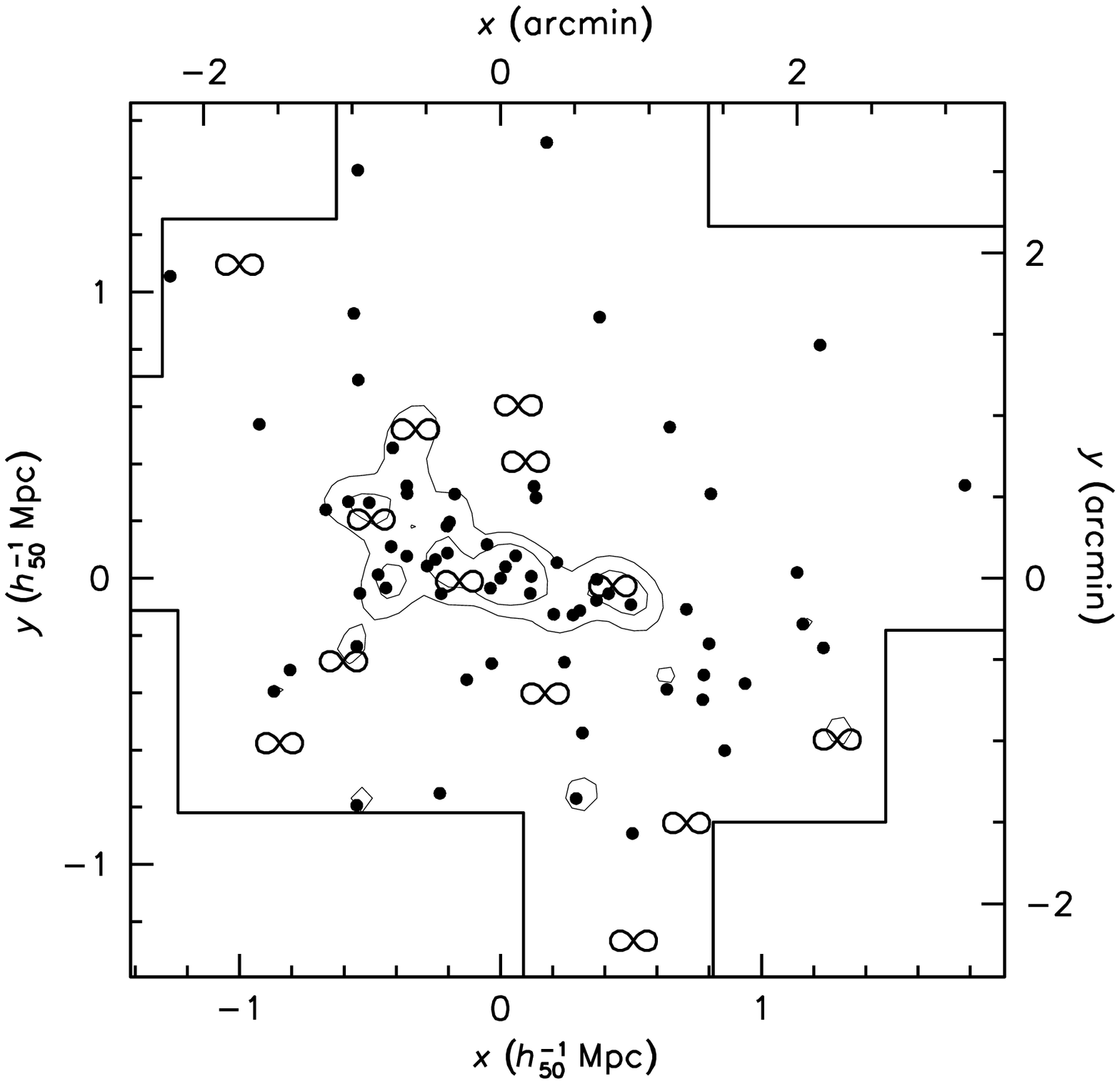}}
\begin{small}
\figcaption{
Spatial distribution of confirmed cluster members in \1054.
Iso-density contours indicate the distribution of red galaxies
in the HST data.
Note the filamentary galaxy distribution.
Mergers are indicated by $\infty$ symbols. The mergers 
are preferentially found in the outskirts of the cluster, indicating
they probably reside in cold infalling clumps.
\label{spatial.plot}}
\end{small}
\end{center}}

Furthermore, the cluster itself is irregular and elongated, as indicated
by the iso-density contours in Fig.\ \ref{spatial.plot} and the X-ray
distribution (Donahue et al.\ 1998). Our data suggest that the
cluster consists of three clumps at the same radial velocity. We
infer that the mergers are possible because the cluster is viewed
before final virialization, and hence before stripping of the halos of
infalling subclumps.

\section{Assembly time versus star formation epoch}

In most cases the
merging galaxies are bulge-dominated, red, and have no detected
\3727{} emission. Hence, they are mergers between ellipticals, S0s or
Sas.  The most striking examples of these gas poor
mergers are galaxies 997 and 1163.  In contrast to mergers in the nearby
field (Liu \& Kennicutt 1995), the star formation rate is very
low, although several galaxies have enhanced Balmer lines indicating a
modest recent star burst.
A major puzzle is how and when the progenitors of the
mergers lost their gas.

The ages of the mergers relative to the other cluster galaxies can be
estimated from their location in the color-magnitude diagram (Fig.\
\ref{cm.plot}).  Mergers
are indicated with $\infty$ symbols.  The median color of the
mergers is modestly bluer (by $\approx 0.07$ magnitudes in $U-B$) than
the CM relation defined by the early-type galaxies, indicating their
luminosity weighted ages are $\approx 40$\,\% lower (Worthey 1994).
Assuming that galaxies on the CM relation formed their stars at $z \geq 3$
(van Dokkum et al.\ 1998b), the stars in the mergers have a luminosity
weighted mean formation redshift $z_* \gtrsim 1.7$.  We conclude that
the mean stellar ages of present-day cluster ellipticals are much larger
than the ages of their last major mergers. We note in passing
that only three of the mergers satisfy the
Butcher \& Oemler (1978) criteria for a blue galaxy.
Assuming $\Delta(U-B) = 1.4 \Delta(B-V)$ (Worthey 1994)
we find the blue fraction in \1054{} is $\sim 20$\,\%; $\sim 75$\,\%
of the blue galaxies are spirals.

\vbox{
\begin{center}
\leavevmode
\hbox{%
\epsfxsize=8cm
\epsffile{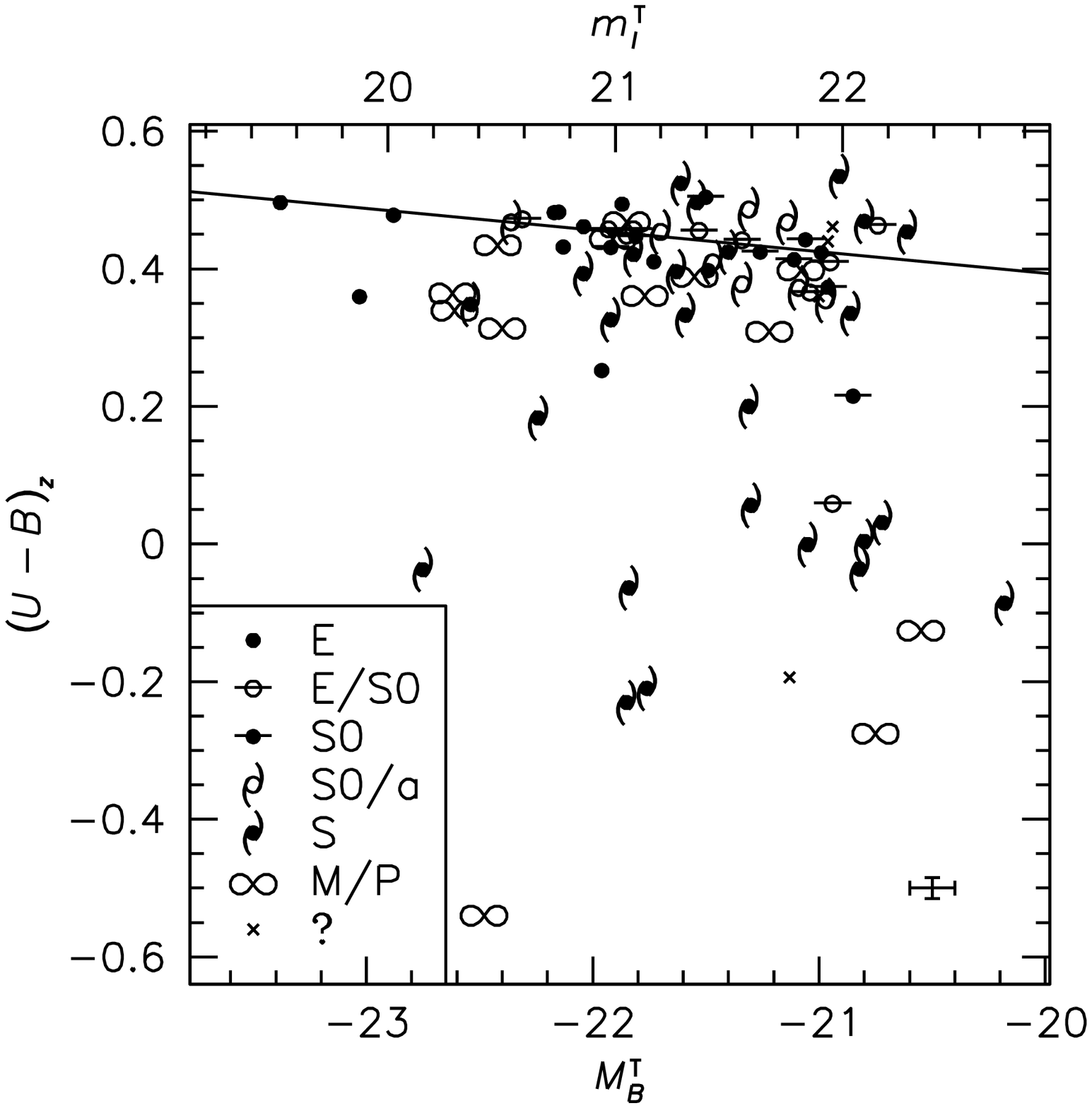}}
\begin{small}
\figcaption{
Color-magnitude relation of spectroscopically confirmed members of
\1054. Restframe $U-B$ colors were derived from observed F606W$-$F814W
colors. Mergers are only $\approx 0.07$ magnitudes bluer than the CM
relation defined by the early-type galaxies (solid line).
\label{cm.plot}}
\end{small}
\end{center}}

We test whether the colors of the mergers are consistent with
measurements of the scatter in the CM relation at $z<1$. The intrinsic
scatter in the CM relation of the existing Es and E/S0s at $z=0.83$ is
low at $0.027 \pm 0.013$ in restframe $U-B$, consistent with the results
of Stanford et al.\ (1998) for other high redshift
clusters.  The scatter in the
combined sample of ellipticals and mergers (i.e., the sample of {\em
future} ellipticals) is much higher, at $0.054 \pm 0.011$. This scatter
will decrease at later times, because fractional age differences
between galaxies will be smaller (e.g., van Dokkum et al.\ 1998a).
We have evolved the CM relation forward in time using the simple models
presented in van Dokkum et al.\ (1998a), and find that the $U-B$ scatter
in the CM relation of ellipticals and mergers will be $\approx 0.035$ at
$z=0.5$, and only $\approx 0.015$ at $z=0$. These numbers are consistent
with the observed scatter in the CM relation at low and intermediate
redshift (Bower, Lucey, \& Ellis 1992; Ellis et al.\ 1997; Stanford et
al.\ 1998; van Dokkum et al.\ 1998a).

\section{Discussion}

The large number of mergers in \1054{} implies strong evolution in the
merger fraction from $z=0$ to $z=0.83$. This is illustrated in Fig.\
\ref{evorate.plot}, which shows the evolution of the merger fraction in
rich clusters with redshift. Solid symbols are CL\,1358+62 at $z=0.33$
and \1054{} at   $z=0.83$.  We determined the merger fraction in
CL\,1358+62 from a sample of 194 confirmed cluster
members imaged with  WFPC2 (van Dokkum et al.\ 1998a). The merger
fraction in this cluster can be compared directly to that in \1054{},
since the sample selection, field size (in Mpc) and classification
method are identical. Merger fractions for other rich clusters were
estimated from the literature; in order of increasing redshift from
Dressler (1980), LH88, C98, LPM92, and D94. These merger fractions are
based on ground based imaging of blue galaxies (LH88, LPM92) or visual
classifications of HST images (D94, C98). The evolution of the merger
rate can be parameterized by $f \propto (1+z)^m$. We find $m =
6.0 \pm 2.0$. The best fit is indicated with the solid line.

\vbox{
\begin{center}
\leavevmode
\hbox{%
\epsfxsize=8cm
\epsffile{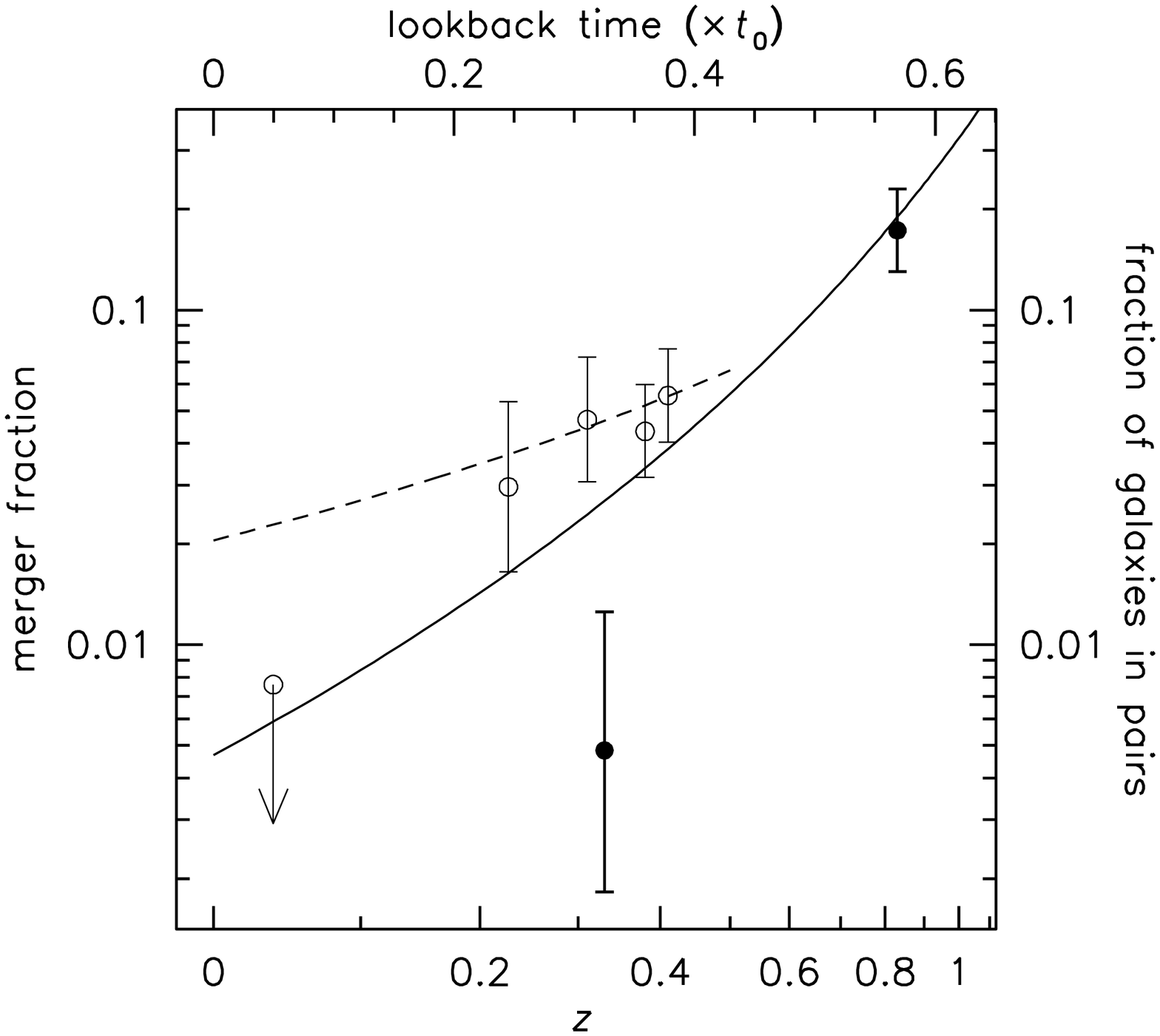}}
\begin{small}
\figcaption{
Evolution of the merger fraction in clusters.
The solid symbols are the clusters
CL1358+62 at $z=0.33$ and \1054{} at $z=0.83$ from our study.
Literature studies of rich clusters are indicated by open symbols.
The solid line is a fit to the
cluster data. The broken line is a fit
to the fraction of field galaxies in close pairs
from Patton et al.\ (1997).
The merger fraction evolves rapidly in clusters and in the field,
possibly even stronger in clusters.
\label{evorate.plot}}
\end{small}
\end{center}}

It is interesting to compare the evolution of the merger fraction in
clusters to that in the field. The broken line in Fig.\
\ref{evorate.plot} is a fit to the fraction of field galaxies in close
pairs, from Patton et al.\ (1997). A direct comparison with the cluster
data is difficult because the latter are based on visual
classifications, which include both close pairs and merger remnants.  We
note, however, that these fractions are comparable in the outer parts of
\1054{}. Although the normalisation is uncertain, the {\em rate} of
evolution can be compared directly. Both the cluster environment and the
field show strong evolution. The evolution in clusters might even be
stronger than in the field. This result is mainly driven by the very low
merger fraction in low redshift clusters. The rapid evolution in the
cluster environment could be due to an enhanced accretion rate onto
clusters at high redshift, and/or enhanced merging during the collapse
of massive clusters. It will be interesting to see whether the merger
fraction correlates with the dynamical state of clusters.

The increase with redshift of the merger rate of massive galaxies is in
qualitative agreement with predictions from hierarchical galaxy
formation models (e.g., Kauffmann 1996, Baugh et al.\ 1996), although it
is a challenge to explain both the recent assembly of massive
early-types and the early formation of their stars. The presence of the
mergers in \1054{} is direct evidence against formation of massive
ellipticals in a ``monolithic'' collapse at very high redshift.

Assuming merging does not alter the shape of the mass function, the
mergers cause an increase in $M_*$ of $\approx 15$\,\%. This can be
contrasted to the strong evolution in the number density of massive
field galaxies inferred by Kauffmann \& Charlot (1998) from the lack of
luminous $K$-band selected galaxies at $z>1$. Although this result is
still uncertain, this may imply that massive galaxies in the
field were assembled more recently than those in clusters.

This study demonstrates that large field studies with HST, in
combination with deep spectroscopy from the ground, can show directly
how galaxy formation proceeded. The large field was essential, since the
mergers are preferentially located in the outskirts of the cluster.
Similar observations of high redshift clusters and the field
would be valuable to test whether the results for
\1054{} are typical.

\acknowledgements{
We thank the referee for constructive and valuable comments.
Support from the University of
Groningen, the University of Leiden,
the Leids Kerkhoven-Bosscha Fonds, and
STScI grant GO07372.01-96A
is gratefully acknowledged.}

\vspace{1cm}
{\noindent \Large NOTE: Figure 1 is available from\vspace{0.2cm}\\
http://www.astro.rug.nl/\~{}dokkum/preprints/merger\_fig1.eps.gz\\
(gzipped postscript)\vspace{0.2cm}\\
http://www.astro.rug.nl/\~{}dokkum/preprints/merger\_fig1.gif\\
(gif)}

\setcounter{figure}{0}

\begin{small}
\figcaption{\label{16bright.plot}
(a) The sixteen most luminous galaxies in the cluster, ordered by total
F814W magnitude (shown in the lower right corner of each image).
Total magnitudes of mergers were calculated from the sum of the
luminosities of both merging galaxies.
Of each galaxy a color image and the F814W
image are shown. The size of each image is $5\farcs9 \times 5\farcs9$
($56 \times 56$\,\h50min\,kpc); the pixel size is
$0\farcs07$. We find five mergers
among the most luminous sixteen galaxies. (b) The eight fainter
galaxies classified as ``merger/peculiar''.
}
\end{small}


\begin{references}
 \reference{}	Barnes, J. E. 1998, in After the Dark Ages: When Galaxies Were
	Young, 9th October Astrophysics Conference, University of
	Maryland, in press (astro-ph/9811242)
 \reference{}	Baugh, C. M., Cole, S., \& Frenk, C. S. 1996,
	\mnras, 283, 1361
 \reference{}	Bower, R. G., Lucey, J. R., \& Ellis, R. S. 1992,
	\mnras, 254, 601
 \reference{}	Butcher, H., \& Oemler, A. 1978, \apj, 219, 18
 \reference{}	Couch, W. J., Barger, A. J., Smail, I., Ellis, R. S.,
	Sharples, R. M. 1998, \apj, 497, 188 [C98]
 \reference{}	Donahue, M., Voit, G. M., Gioia, I., Lupino, G., Hughes,
	J. P., Stocke, J. T. 1998, \apj, 502, 550
 \reference{}	Dressler, A. 1980, \apj, 236, 351
 \reference{}	Dressler, A., Oemler, A., Jr., Sparks, W. B., \&
	Lucas, R. A. 1994, \apj, 435, L23 [D94]
 \reference{}	Dressler, A., Oemler, A., Jr., Couch, W. J., Smail, I.,
	Ellis, R. S., Barger, A., Butcher, H., Poggianti, B. M.,
	\& Sharples, R. M. 1997, \apj, 490, 577
 \reference{}	Ellis, R. S., Smail, I., Dressler, A., Couch, W. J.,
	Oemler, A., Jr., Butcher, H., \& Sharples, R. M. 1997, \apj, 483, 582
 \reference{}	J\o{}rgensen, I., \& Franx, M. 1994, \apj, 433, 553
 \reference{}	Kauffmann, G. 1996, \mnras, 281, 487
 \reference{}	Kauffmann, G., \& Charlot, S. 1998, \mnras, 297, L23
 \reference{}	Lavery, R. J., \& Henry, J. P. 1988, \apj, 330, 596 [LH]
 \reference{}	Lavery, R. J., Pierce, M. J., \& McClure, R. D. [LPM]
	1992, \aj, 104, 2067
 \reference{}	Liu, C. T., \& Kennicutt, R. C., Jr. 1995, \apj, 450, 547
 \reference{}	Lubin, L. M., Postman, M., Oke, J. B., Ratnatunga,
	K. U., Gunn, J. E., Hoessel, J. G., \& Schneider, D. P. 1998,
	\aj, 116, 584
 \reference{}	Merritt, D. 1984, \apj, 276, 26
 \reference{}	Oke, J. B., et al. 1995, \pasp, 107, 375
 \reference{}	Patton, D. R., Pritchet, C. J., Yee, H. K. C.,
	Ellingson, E., \& Carlberg, R. G. 1997, \apj, 475, 29
 \reference{}	Rix, H-W. R., White, S. D. M. 1989, \mnras, 240, 941
 \reference{}	Roos, N., \& Aarseth, S. J. 1982, A\&{}A, 114, 41
 \reference{}	Searle, L., Sargent, W. L. W., \& Bagnuolo W. G. 1973,
	\apj, 179, 427  
 \reference{}	Stanford, S. A., Eisenhardt, P. R., \& Dickinson, M.
	1998, \apj, 492, 461
 \reference{}	Toomre, A., \& Toomre, J. 1972, \apj, 178, 623
 \reference{}	Tran, K-V. H., Kelson, D. D., van Dokkum, P. G., Franx,
	M., Illingworth, G. D., \& Magee, D. 1999, \apj, in press
 \reference{}	van Dokkum, P. G. 1999, PhD thesis, Groningen University
 \reference{}	van Dokkum, P. G., Franx, M., Kelson, D. D.,
	Illingworth, G. D. I., Fisher, D., \& Fabricant, D. 1998a, \apj,
	500, 714
 \reference{}	van Dokkum, P. G., Franx, M., Kelson, D. D.,
	\& Illingworth, G. D. 1998b, \apj, 504, L17
\reference{}	Worthey, G. 1994, \apjs, 95, 107
\end{references}
\end{document}